\documentclass{ifacconf}

\usepackage{graphicx}      
\usepackage{natbib}        
\usepackage{mathrsfs}
\usepackage{amsmath,amsfonts}
\usepackage{xfrac}
\usepackage{color,soul}
\usepackage{bbold}
\usepackage{microtype}
\usepackage{booktabs}
\usepackage{marginnote}
\newtheorem{definition}{Definition}[section]
\DeclareMathOperator*{\argmin}{arg\,min}

\newif\ifmargincomments 
\margincommentstrue

\ifmargincomments

\else

\fi
\begin{document}
\begin{frontmatter}

\title{A Data-driven Pricing Scheme for Optimal Routing through Artificial Currencies} 

\author[First]{David\,van\,de\,Sanden} 
\author[Second]{Maarten\,Schoukens}
\author[Third]{Mauro\,Salazar}

\address[First]{Eindhoven University of Technology, (e-mail: davidvandesanden@live.nl)}
\address[Second]{Department of Electrical Engineering,
Eindhoven University of Technology, The Netherlands, (e-mail:
m.schoukens@tue.nl)}
\address[Third]{Department of Mechanical Engineering,
Eindhoven University of Technology, The Netherlands, (e-mail:
m.r.u.salazar@tue.nl)}

\begin{abstract}
Mobility systems often suffer from a high price of anarchy due to the uncontrolled behavior of selfish users. This may result in societal costs that are significantly higher compared to what could be achieved by a centralized system-optimal controller. Monetary tolling schemes can effectively align the behavior of selfish users with the system-optimum. Yet, they inevitably discriminate the population in terms of income. Artificial currencies were recently presented as an effective alternative that can achieve the same performance, whilst guaranteeing fairness among the population. However, those studies were based on behavioral models that may differ from practical implementations.
This paper presents a data-driven approach to automatically adapt artificial-currency tolls within repetitive-game settings. 
We first consider a parallel-arc setting whereby users commute on a daily basis from an individual origin to an individual destination, choosing a route in exchange of an artificial-currency price or reward, while accounting for the impact of the choices of the other users on travel discomfort.
Second, we devise a model-based reinforcement learning controller that autonomously learns the optimal pricing policy by interacting with the proposed framework considering the \textit{closeness} of the observed aggregate flows to a desired system-optimal distribution as a reward function.
Our numerical results show that the proposed data-driven pricing scheme can effectively align the users' flows with the system optimum, significantly reducing the societal costs with respect to the uncontrolled flows (by about 15\% and 25\% depending on the scenario), and respond to environmental changes in a robust and efficient manner.

\end{abstract}

\end{frontmatter}

\section{Introduction}
Worldwide urbanization, population growth, and motorization are currently leading to global surges of congestion~\citep{texasmobilityreport}. As a result, central authorities such as governments are looking for ways to minimize the discomfort of their mobility users as well as to reduce pollution and emissions. Although collaboration between groups of users can lead to a significant gain in the overall efficiency of a transportation network~\citep{IAMoD,zhang2015models}, the selfish nature of humans often prevents the system from performing in a system-optimal fashion on the ``macro" level~\citep{SelfishRouting}.
In this context, the challenge for central operators is to align the routing choices of selfish users with the system optimum, whilst still allowing for some degree of individual autonomy.
\begin{figure}[t]
    \centering
    \includegraphics[width=\columnwidth]{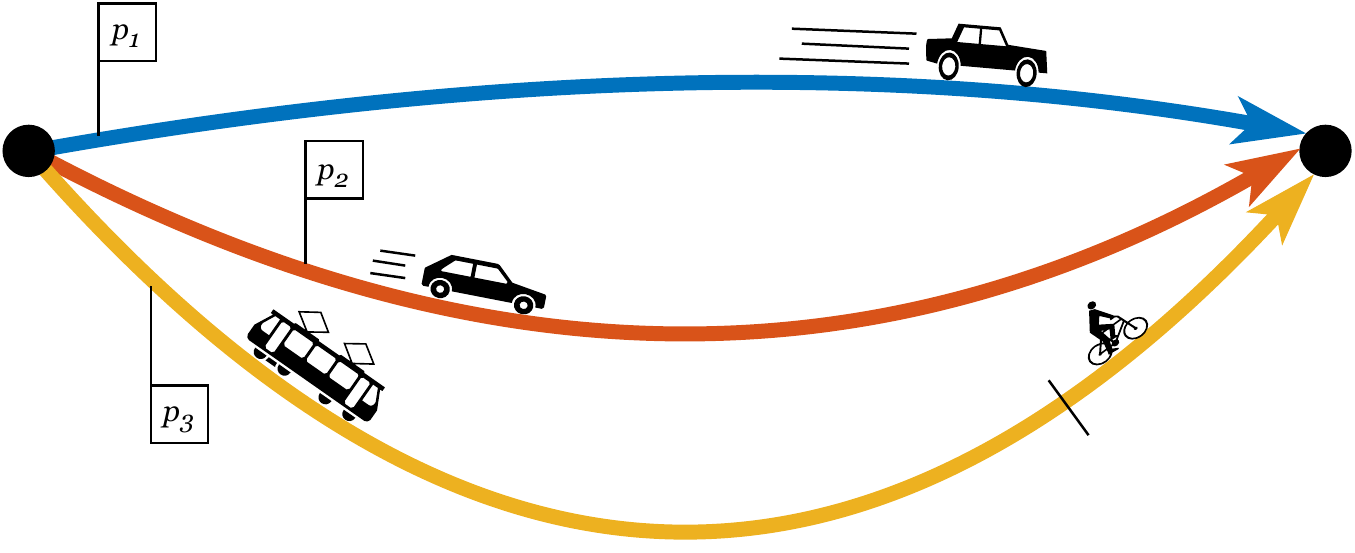}
    \caption{Mesoscopic transportation network consisting of one origin node, one destination node, and a set of three available arcs for commuters to choose from. Associated to each route is a price in Karma $p_j \in \mathbb{R}$ set by the proposed data-driven toll-pricing controller.}
    \label{fig:RoadNetwork}
\end{figure}
To this end, monetary tolls have often been considered an effective solution to achieve such an alignment~\citep{pigou1920economics,morrison1986survey,bergendorffTolling}. Yet they suffer from a fundamental drawback: They are inherently financially discriminatory.
In this paper, we try to solve this issue leveraging artificial currencies instead.
In line with its first introduction as \textit{Karma} in~\cite{Censi2019TodayMT} and its application to mobility systems in~\cite{SALAZAR2021}, this currency can be spent or gained only by traveling, but since it can neither be bought nor exchanged it loses any monetary value and financial discriminatory power.
The aforementioned and subsequent studies showed the potential of artificial currencies by adopting optimization-based behavioral models to describe presumably rational users interacting with artificial-currency schemes which, however, have not been validated yet.
In this paper, we circumvent the lack of validated models by proposing a data-driven tolling scheme that automatically learns and induces the prices optimal to align the behavior of selfish users with the desired system optimum by directly reacting to the observed aggregate flows and \textit{Karma} levels.
\subsubsection*{Related work:}
This paper pertains to two research streams: artificial currencies for resource allocation and data-driven tolling schemes.
Recently, artificial currencies have been successfully integrated within bidding mechanisms~\citep{prendergast2016allocation,gorokh2021monetary,Censi2019TodayMT,ezzat_censi_karma}.
Yet such bidding mechanisms may result in decision fatigue and users' dissatisfaction due to a lack of an upfront decision (when bidding, users do not know whether they will be granted their choice).
Against this backdrop, judiciously-chosen fixed prices that are known upfront to the user proved to guarantee asymptotic convergence to the desired system optimum within two-arc repetitive-game settings~\citep{SALAZAR2021}.
Yet all existing methods are based on pre-defined analytical behavioral models lacking practical validations. Hence, they may result in non-functioning practical implementations, a limitation that we propose to solve by leveraging data-driven methods.

Data-driven pricing schemes have lately been investigated for conventional monetary tolls. Due to an assumed lack of a behavioral routing model, as well as a lack of labeled samples required for supervised learning, most current research streams select reinforcement learning as the preferred type of machine learning. For example, in \cite{chen2018dyetc} a policy-gradient (PG) methodology is devised for the dynamic pricing of electric tolls of a graph-based transportation network. Concurrently, \cite{pandey2020deep} propose a neural network-based PG method to develop a pricing scheme for High-occupancy/toll lanes. In general, PG is considered more effective in the pricing scheme development compared to, e.g., value-based methods for its superior efficiency and capabilities of solving large scale Markov decision processes. However, both the aforementioned works only focus on devising pricing schemes for monetary tolls, which inherently display financial discrimination. Additionally, both aforementioned approaches require in the order of magnitude of $10^5$ interactions with an environment before arriving at a (near-)optimal policy, thereby limiting the application to simulation environments only. We significantly differ from this approach by introducing a data-efficient model-based RL method suitable for real-world implementation.

In conclusion, to the best of the authors' knowledge, there are currently no data-driven tolling schemes based on artificial currencies for mobility systems. 
\subsubsection*{Statement of contributions:}
This paper introduces a data-driven controller that learns a state-dependent toll-pricing policy through direct interaction with a dynamical environment. Starting from the mesoscopic repeated game framework and individual routing model of \cite{SALAZAR2021}, we first present a model-based RL controller for the autonomous devisal of a near-optimal Karma-based pricing scheme. For this controller, prices placed on the arcs are kept constant on individual days, but are allowed to vary day-to-day. Second, we present a numerical episode of learning for a transportation network with three arcs. It is important to emphasize the absence of any assumed knowledge on the routing behavior of users to the controller, thereby decoupling the modeling accuracy of routing behavior and overall system performance. Finally, we test the robustness of our proposed controller by implementation within a non-stationary environment. 

\subsubsection*{Organization}
The remainder of this work is structured as follows: We present the general framework and problem in which the data-driven tolling scheme is formulated in Section \ref{sec:Problem}. In Section \ref{sec:Methodology} we present the methodology of the data-driven toll-pricing controller. Thereafter we discuss any limitations to our current approach. We subsequently present the numerical results obtained by training a data-driven controller through interaction with a mesoscopic routing model in Section \ref{sec:NumResults}. In Section \ref{sec:Conclusion} we discuss our findings and make recommendations for further research.
\section{Problem Formulation}
\label{sec:Problem}
This section introduces the routing problem we aim to solve within a repetitive-game setting, and formulates the control objectives.

We consider a transportation network represented by a directed graph. More specifically, we examine the case of a parallel-arc transportation network between a single origin node and a single destination node. Connecting these two nodes are $n\in \mathbb{N}_+$ directed arcs. Fig.~\ref{fig:RoadNetwork} depicts the case where $n=3$. At $t\in \mathbb{N}$ discrete time steps, all network users are offered the choice of traveling via one of the $n$ arcs. This repeated game setup emulates the simulation of a group of users being offered the choice of a set of routes or modes of transportation in a daily commute. Each individual user $i\in\{1,2,\dots,M\}$, whereby $M$ is the total amount of users, decides whether or not to travel a specific arc $j\in \{1,2,\dots,n\}$, indicated by the binary variable $y_{j}^i(t)$, where $y_{j}^{i}(t) = 1$ if person $i$ chooses arc $j$ and 0 otherwise.
Furthermore, at any time $t$ every user $i\in \{1,2,...,M\}$ is limited to travel via one arc, at most:
\begin{align*}
    \mathbb{1}^\top y^{i}(t)\leq 1,
\end{align*}
where $y^i(t)\in \{0,1\}^n$ is a vector of binary indices containing information on which arc user $i$ chooses to travel. Then, we capture the aggregate flows in $x(t)\in\mathbb{R}^n$. Specifically, component $x_j(t)$ of vector $x(t)\in[0,1]^n$ represents the fraction of users choosing arc $j$ at time $t$:
\begin{equation}
\label{eq:trafficflow}
    x_j(t)=\frac{1}{M}\sum_{i=1}^{M} y_{j}^{i}(t). 
\end{equation}
Furthermore, at every time step $t$, a user has a probability $P_\mathrm{home} \in [0,1]$ of staying at home and a probability of commuting $P_\mathrm{go}=1-P_\mathrm{home}$.
Assuming user $i$ can own a non-negative amount of Karma $k^i(t)\in \mathbb{R}_+$, combined with its choice scheme $y^i(t)$, its Karma balance is updated as follows:
\begin{align}
\label{eq:updateKarma}
    k^i(t+1) = k^i(t) - p^\top(t) y^i(t),
\end{align}
with $p(t)\in[p_\mathrm{min},p_\mathrm{max}]^n$ the price in Karma set by the toll-pricing controller on each of the arcs.
Crucially, some arcs can have negative prices, i.e., rewards, such that users will be reimbursed when crossing them. To prevent users from spending indefinite Karma, we oblige users to have positive Karma levels at any points in time, i.e., $k^i(t)\geq 0\;\forall i,t$.

We make a distinction between the personal discomfort an individual user experiences and the societal cost perceived by a central operator, both as a function of the road occupancy $x(t)$. The personal discomfort $d:[0,1]^{n} \xrightarrow{} \mathbb{R}_{+}^{n}$ is the cost an individual would experience on each of the $n$ arcs, if they were to travel at time $t$.
Note that this cost is equal for all users opting to travel arc $j$; for instance, it could represent the travel time of arc $j$ given its current occupancy. The flow-dependent societal cost \mbox{$c:[0,1]^{n} \xrightarrow{} \mathbb{R}_{+}^{n}$} is set by a central operator and can be aligned with minimizing the aggregate perceived discomfort, i.e., $c(x(t))=d(x(t))$, or also be chosen for other aims, e.g., minimizing congestion and emissions. Individual arc-specific components of the personal or societal cost vectors are indicated by $d_j(x_j(t))$ and $c_j(x_j(t))$, respectively.
\subsection{Central Operator's Problem}
Similar to \cite{SALAZAR2021}, we assume the presence of a central authority (e.g., a city council), which needs to give an incentive to its users such that the aggregate traffic flows converge to the minimizer of the total societal cost $C(x)=c(x)^\top x$. Establishing the optimal distribution of mesoscopic traffic flows $x^\star$ is the solution of the following problem:

\begin{prob}[Optimal Mesoscopic Flows]\label{prob:main}
The central operator aims at routing users so that the aggregate route choice
$x$ converges to
\begin{subequations}
\begin{align}
x^\star\in \underset{x\in [0,1]^n}{\arg\min} & \; C(x) \\
\textnormal{s.t.} & \; \mathbb{1}^\top x = P_{\mathrm{go}} \label{eq:constraintProb1}.
\end{align}
\end{subequations}
\end{prob}
\noindent 
Note that all users wishing to commute should be permitted to do so, which is enforced through~\eqref{eq:constraintProb1}. In the same spirit as \cite{SALAZAR2021}, we make the assumption that the optimal solution $x^\star$ is either known or can be computed through present knowledge of the multi-arc transportation system. 

Assuming the optimal distribution of mesoscopic traffic flows $x^\star$ is known, the second challenge a central operator faces is how to devise a Karma pricing policy $\pi:\mathcal{S}\xrightarrow{} \mathbb{R}^n$, whereby $\mathcal{S}$ is the admissible state space, which steers the population of users towards $x^\star$. A fundamental aspect of this problem is the way in which human beings make routing choices in combination with artificial currencies. We assume the central operator to have \textit{no} prior knowledge of the mesoscopic choice behavior of its population due to three reasons: (1) a lack of empirical data for transportation systems with artificial currencies, (2) modeling discrepancies concerning rational economic human behavior \citep{nonrational} and (3) possible regional and temporal differences in mesoscopic choice behavior. Instead, the central operator aims at nudging the aggregate routing towards $x^\star$ by implementing a data-driven toll-pricing controller, through which it strives to find the optimal pricing policy~$\pi^\star$. Furthermore, we intend to learn the optimal toll-pricing policy in as short a time as possible, so as to minimize the total societal cost. 

\begin{prob}[Pricing Problem]\label{prob:pricing}
Given a desired system optimum $x^\star$, the pricing problem consists of finding a parameterized pricing policy $\pi(\theta,s)$, whereby $s\in \mathcal{S}$ is the environment's state, which minimizes the expected distance between $x^\star$ and $x(t)$ over a horizon length $H$:
\begin{equation}
\label{eq:pricing_problem}
\begin{aligned}
J^\pi(\theta) = \sum_{t=0}^H \mathbb{E}_\pi[D(x(t),x^\star)],
\end{aligned}
\end{equation}
where $J^\pi$ is the objective function, $D:\mathcal{S}\xrightarrow{}\mathbb{R}_+$ a distance measure and $\theta$ constituting the parameters of the pricing policy. Note that the road distribution $x(t)$ is affected by the pricing policy $\pi$ through the choice behavior of the population, and hence the control parameters $\theta$.
\end{prob}

\subsection{Simulation of Routing Behavior}
In order to benchmark our data-driven controller, we leverage a simulation environment describing the routing choices of \emph{rational} individuals.
Though not being limited to it, in this work we apply the data-driven toll-pricing controller in the optimization-based analytical model presented in~\cite{SALAZAR2021}.
Thereby, each individual user determines their own optimal route by minimizing the discomfort perceived today with some sensitivity $q$ (e.g., rush), and the discomfort perceived for the rest of a time-horizon $T$ (e.g., a week) with the average of their sensitivity $\bar{q}$, so that their Karma level will never be negative and is above some reference level $k_\mathrm{ref}$ at the end of the planning horizon:

\begin{prob}[Individual Agent's Problem]\label{prob:individualagent}
At time $t$, given the flows $x$ and prices $p$, respectively, a traveling agent with Karma level $k\geq0$ and reference $k_\mathrm{ref}$, and sensitivity $q$ will choose their route as $y^\star$ resulting from
\begin{subequations}
\begin{align}
(y^*,\Bar{y}^\star) \in \argmin_{y\in\mathcal{Y},\Bar{y}\in\mathcal{\Bar{Y}}} & \, q \cdot d(x)^\top y + T \cdot \Bar{q} \cdot d(x)^\top \Bar{y}  \\
s.t. & \, k - p^\top y - T \cdot p^\top \Bar{y} \geq k_\mathrm{ref} \\
     & \, p^\top y \leq k,
\end{align} \label{eq:ind_agent}
\end{subequations} 
\hspace*{-3mm} with $\mathcal{Y}=\{y\in\{0,1\}^n:\mathbb{1}^\top y=1\}$ and $\mathcal{\Bar{Y}}=\{\Bar{y}\in\{0,1\}^n:\mathbb{1}^\top y=1\}$, and $T$ as the individual's planning horizon.
We define the set containing all points $y^\star$ solving \eqref{eq:ind_agent} as $\mathcal{Y}^\star (x,q,k,k_\mathrm{ref})\subseteq \mathcal{Y}$.
Non-traveling agents have $y^\star=\mathbb{0}$.
\end{prob}
To link individual choices to aggregate flows, at every time $t$, we assume the users to reach a Nash equilibrium, whereby no user would be able to improve their objective by unilaterally changing their choice. A more detailed overview can be found in~\cite{SALAZAR2021}.
Finally, we define the uncontrolled aggregate flows resulting from the absence of the tolling scheme as follows:

\begin{definition}[Uncontrolled Road Distribution]
\label{def:xuc}
We define the uncontrolled road distribution $x_{\mathrm{UC}}$ to be the expected road distribution when no Karma pricing scheme is in place, i.e., ${x^{\mathrm{UC}}=\mathbb{E}[x|p=\mathbb{0}]}$.
\end{definition}

\section{Pricing Controller}
\label{sec:Methodology}

Our proposed toll-pricing controller needs to meet two criteria. First, the controller needs to be able to learn its pricing scheme  in a purely data-driven fashion. Second, it needs to have high data efficiency, i.e., learn the optimal policy with minimal interaction or trials. This second criterion comes from an assumed lack of a high-fidelity simulator. Due to this absence, we intend to directly implement our data-driven approach in a real-world situation. Considering that our system includes real human beings, implementing poor policies for longer periods of time can only be considered as undesirable or even unethical. A substantial advantage of learning the pricing scheme directly from real-world data is that any modeling errors, which could be present in simulators, are circumvented. For the two aforementioned criteria, we propose to base our controller on the probabilistic inference for learning and control (PILCO) method presented in~\cite{deisenroth2011pilco}. 

PILCO can be classified as a \textit{model-internalization} RL-algorithm, as it revolves around the development and employment of a stochastic dynamical model via interactions with a dynamical environment.
Given this learned model of the environment dynamics, PILCO subsequently makes hypothetical trials and improves its policy offline, making it highly data-efficient throughout training \citep{deisenroth2011pilco}.
The iterative pipeline of PILCO is composed of three subsequent stages: 1) the estimation of a probabilistic dynamical model, 2) an evaluation of the policy, and 3) a policy update. The updated policy $\Bar{\pi}$ is applied to the system for $t_\mathrm{update}$ time steps before the recorded data is used again for learning the probabilistic dynamical model.

\subsubsection*{States:}
The considered set of states $\mathcal{S}$ in this research consists of two elements: 1) the aggregate mesoscopic traffic flow $x(t)$ at the current time $t$ and 2) the Karma distribution of the complete population $K(t)= [k^1(t), k^2(t), \dots, k^M(t)]$. For privacy reasons, we only assume the mean $\Bar{K}$ and standard deviation $\sigma_K$ of the Karma distribution of the population to be known to the data-driven controller. Formally, we define the observable state $s$ as the 3-tuple $\langle x,\Bar{K},\sigma_K \rangle$.

\subsubsection*{Actions:}
The controller's action set $\mathcal{A}$ consists of a bounded, continuous, Karma-based set of prices it can set for each of the arcs. Bounding the action space imposes no limitations on the optimality of the problem, as \cite{SALAZAR2021} show that the magnitude of the price does not influence the optimal road distribution once converged. However, the transient behavior of mesoscopic traffic flows can be strongly influenced by changing the magnitude of the prices. This is why a pricing policy, with fixed prices on individual days, but possible day-to-day fluctuations, opens significant possibilities for further reducing the accumulated perceived societal cost.

\subsubsection{Dynamical Model:}
A Gaussian Process (GP) is used for estimating a stochastic dynamical model of the environment at hand. A GP is a non-parametric method for modeling environment dynamics that revolves around a kernel- or covariance function. In particular, throughout this study a squared-exponential covariance function is used. In the GP, we use tuples $\langle s_t, p(t) \rangle$, with $s_t \in \mathcal{S}$ the state at time $t$, as training inputs and differences \mbox{$\Delta_t = s_{t+1} - s_t$} as training targets. The aim of this stochastic dynamical model is to be able to make offline hypothetical trials on which the current policy can be improved. By incorporating a form of stochasticity, the policy accounts for any uncertainties present in the current estimated dynamical model. For an extensive overview of GPs in machine learning, we refer the reader to \cite{williams2006gaussian}.

\subsubsection{Policy Evaluation:}
To evaluate and update the objective function $J^\pi$ based on the current policy $\pi$, long-term predictions are made of future state distributions $f(s_0),\dots,f(s_T)$. These future state distributions are devised through cascading one-step predictions. However, calculating the subsequent state prediction $f(\Delta_t)$ entails solving an analytically intractable integral
\begin{equation}
    f\left({\Delta}_{t}\right)=\iint f\left(g\left(\tilde{{s}}_{t}\right) |\, \tilde{{s}}_{t}\right) f\left(\tilde{{s}}_{t}\right) \mathrm{d} g \mathrm{~d} \tilde{{s}}_{t},
\end{equation}
with $f(\tilde{{s}}_{t})=f(s_t,a_t)$ a joint probability distribution between states $s_t$ and actions $a_t$. To alleviate this intractability, the predictive distribution is approximated by a Gaussian distribution by means of moment matching. Through this approximation, future state predictions and hence the expected future costs can be computed.

\subsubsection{Policy Update:}
PILCO is a policy search method that uses the gradient of the cost function with respect to the parameters of its policy $\frac{\partial J^\pi}{\partial\theta}$, to update and improve its policy. By using the chain-rule, analytic derivations of this derivative are found and implemented in a gradient-based non-convex optimization algorithm to look for the set of parameters that minimize the formulated objective function. Once convergence is achieved, the updated policy is implemented, new data is observed, and subsequently a new learning iteration starts. For a more detailed overview of PILCO, we refer the reader to \cite{deisenroth2011pilco,deisenroth2013gaussian}.

\subsection{Pricing Policy}
The controller policy $\pi$ within PILCO is formulated as a state-feedback controller. In this parameterized policy, the optimal parameters $\theta^\star$ are learned through the previously mentioned iterative pipeline. For the task of aligning and stabilizing the aggregate routing $x(t)$ at the optimal aggregate road distribution $x^\star$, we opt for a linear state-feedback controller
\begin{equation}
    \tilde{\pi}(s)= As + b,
\end{equation}
with $\tilde{\pi}$ a preliminary linear state-feedback controller, and $A$ and $b$ components of the parameter vector $\theta$. As the magnitude of the prices does not have an effect on the stationary distribution \citep{SALAZAR2021}, we opt to limit the magnitude of the available prices. This input constraint is enforced through a squashing function
\begin{equation}
\label{eq:squash}
    \zeta(s) = \frac{9}{8}\sin (s) + \frac{1}{8}\sin(3s) \;\; \in [-1,1],
\end{equation}
which is the third-order Fourier series expansion of a trapezoidal wave, normalized to be within the interval $[-1,1]$. The combined policy is then computed by passing the preliminary policy through the squashing function before scaling it with the maximum amplitude of the signal:
\begin{equation}
    \pi(s)= \max(|p_{\mathrm{min}}|,|p_{\mathrm{max}}|) \cdot \zeta(\tilde{\pi}(s)).
\label{eq:squashing_policy}
\end{equation}
Subsequently, the control distribution $f(a_t)$ can be calculated analytically for Gaussian distributed states and used for the long-term predictions of the policy evaluation.

Finally, we give every user the possibility to make their commute whenever they seek to do so. Even for a bounded policy, scenarios can be formulated where all arcs have positive prices, resulting in users being prevented from traveling even though they were prepared to take the least comfortable arc. Hence, we restrict the prices to $p_1\geq 0$, and $p_n \leq 0$ for the most and least comfortable arc, respectively. We enforce this by parsing the proposed prices $\tilde{p}_1$ and $\tilde{p}_n$ through a softplus function: $p_1(t)=\log (1 + \exp{(\tilde{p}_1(t))})$ and $p_n(t)=-\log (1 + \exp{(-\tilde{p}_n(t))})$. Note that this transformation is included within the environment and hence not included in the control distribution $f(a_t)$.

\subsection{Initial Pseudo-Random Policy}
Originally, PILCO requires an initial period in which the sole purpose is gathering informative data. \cite{deisenroth2011pilco} propose an initial window in which a random policy is used. As we assume the local authority to already have some prior knowledge of the transportation network, we aim at improving this initial random policy. Specifically, we assume a general understanding of which arc is predominantly seen as the most comfortable or desirable arc. We can capture this information by assigning the fastest or most comfortable arc with the highest price, the second fastest arc with the second highest price, etc. More specifically, we first draw samples from a uniform distribution between $p_\mathrm{min}$ and $p_\mathrm{max}$ and subsequently set the prices at time $t$ as the sorted samples in descending order:
\begin{equation}
    p_j(t) \sim \mathcal{U}(p_{\mathrm{min}},p_{\mathrm{max}}), \; \forall t\leq t_\mathrm{init} \forall j\in \{1,2,\dots ,n\}
\end{equation}
\begin{equation}
    p_1(t) \geq p_2(t) \geq ... \geq p_n(t),
\end{equation}
where $t_\mathrm{init}\in \mathbb{N}$ is the length of time in which the initial pseudo-random policy is implemented.

\subsection{Objective and Cost Function}
PILCO has initially been set up as a method for tracking predefined trajectories within robotic systems. Therefore, the most effective cost functions revolve around some form of distance function with respect to a predefined target state $s^\star$. Furthermore, we formulate the distance function $D$ of~\eqref{eq:pricing_problem} as a saturated cost function:
\begin{equation}
    D(s(t)) = 1 - \exp (-\frac{1}{2\sigma_c^2} (s(t)-s^\star)^\top W (s(t)-s^\star)),
    \label{eq:saturated_distance}
\end{equation}
where $\sigma_c^2$ is the width of the saturating cost function and $W$ is a diagonal matrix with entries either one or zero. It is worth calling to mind that $D(s) \in [0,1]$ has the shape of an inverted normal distribution, which indirectly encourages exploration of the system through a near constant cost at states that are far from the target. In our framework, we only desire to control the aggregate road distribution $x(t)$, and have no interest in controlling any of the other states. Therefore the entries of $W$ that correspond to $\Bar{K}$ and $\sigma_K$ are set to zero.

\subsection{Discussion}
A few comments are in order.
First, we assume the central operator to have knowledge of the optimal mesoscopic traffic flow $x^\star$. However, scenarios can be devised where this knowledge is not available. In these cases, the cost function of the data-driven controller needs to be adjusted accordingly. We leave devising such cost functions to further research. It is worth noting that a direct implementation of the aggregate societal cost $C(x(t))$ will not lead to satisfactory results, as the controller will devise a policy that forces all users to stay at home, i.e., $x(t)=\mathbb{0}$. 
Second, the data-driven controller currently has a hard constraint on having at least one arc with a negative price. However, this constraint is enforced as part of the environment, and is therefore not taken into account throughout the optimization. RL-methods such as \cite{kamthe2018data} can incorporate input constraints directly. Yet incorporating these constraints might result in a slight loss of performance in terms of convergence speed.
Third, more complex controllers such as radial-basis-function networks could increase performance in the (highly) nonlinear regions of the system dynamics, but since our aim is predominantly to stabilize the aggregate road distribution at the optimum, we leave it to future research to include these types of controllers.
Fourth, our method gives equal weight to all collected data samples. One could consider giving the latest data samples a larger importance, i.e., implementing a forgetting factor, so as to increase performance. Possible methods for updating the GP model can be found in \cite{rottmann2010learning}.
\section{Numerical Results}
\label{sec:NumResults}
This section presents the numerical results achieved by applying the proposed controller on the routing environment from \cite{SALAZAR2021}. More specifically, we display a single episode of learning consisting of an initial pseudo-random policy, followed by an iterative updating process. Furthermore, we display the advantages of continuously updating the pricing policy by applying the controller to a non-stationary dynamical environment. For all displayed scenarios, we represent the personal discomfort $d$ as the travel time model from the \cite{united1964traffic}:
\begin{align}
    d_j\left(x_{j}\right)=d_{j}^0 \cdot\left(1+\alpha \cdot\left(\frac{x_{j}}{ \kappa_{j}}\right)^{\beta}\right),
\end{align}
with a free-flowing time $d_{j}^0$, road capacity parameter $\kappa_j$, and $\alpha$ and $\beta$ further road-specific BPR parameters. Additionally, we set $p_{\mathrm{max}}=-p_{\mathrm{min}}=20$.

\subsection{Pricing Scheme Learning Process}
In a three-arc transportation network ($n=3$), we set $d^0=[1,1.5,1.5]^\top$, $\kappa=[0.5,0.5,0.9]^\top$, $\alpha=0.15$ and $\beta=4$. For the reference Karma values we draw samples from a Gaussian distribution $k_\mathrm{ref}^i \sim \mathcal{N}(50,10)$. Furthermore, we set the number of users to $M=500$, $k^i(0) \sim \mathcal{U}(50,100)$, $t_{\mathrm{init}}=10$ and $t_{\mathrm{update}}=5$.
We sample users' daily sensitivities from an exponential distribution defined on $\mathbb{R_+}$.
The optimal mesoscopic flows are $x^\star=[0.45,0.18,0.32]^\top$ and the uncontrolled mesoscopic flows $x^\mathrm{UC}=[0.68,0.10,0.17]^\top$. 
\begin{figure}[tb]
    \centering
    \includegraphics[width=\columnwidth]{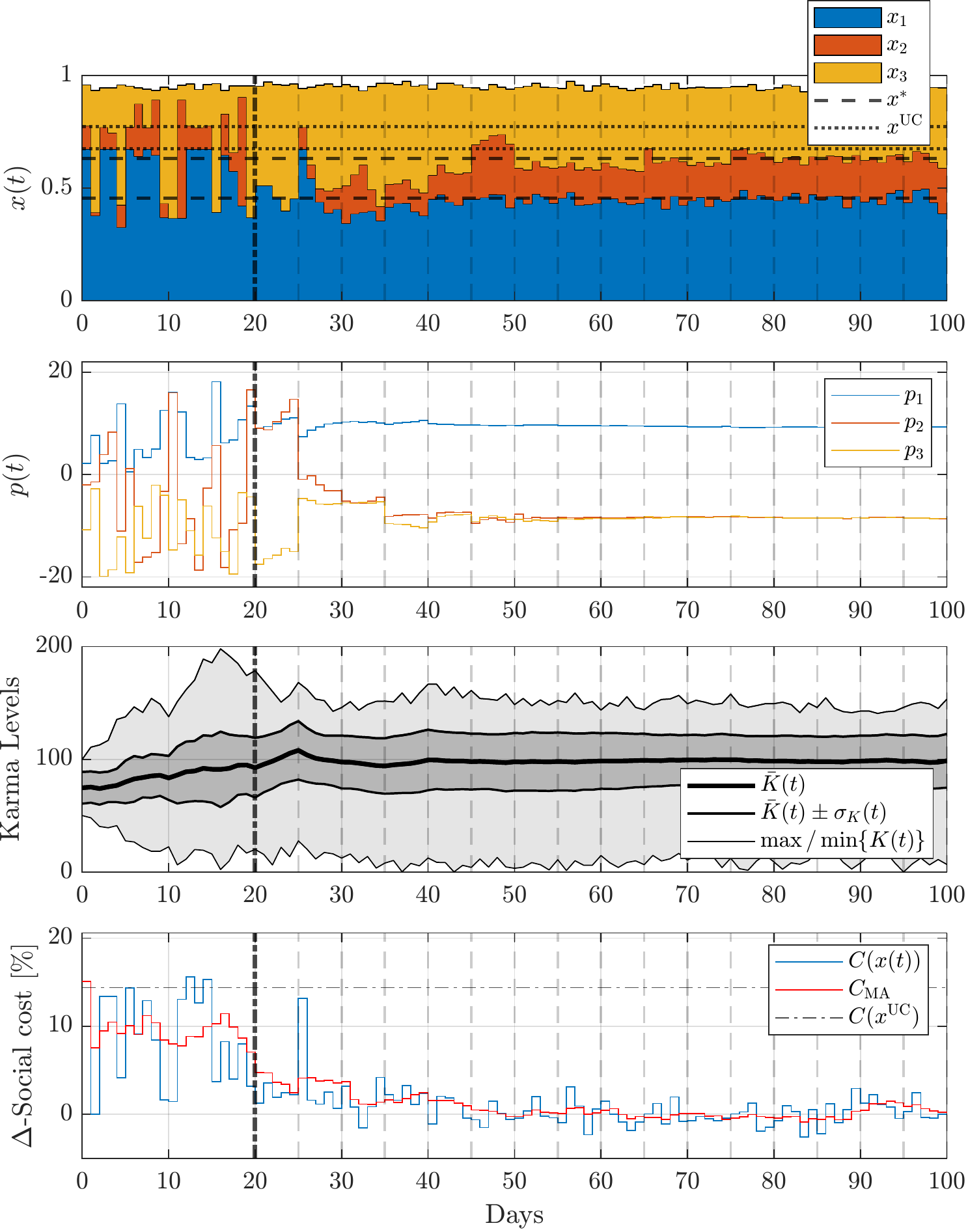}
    \caption{Mesoscopic flows, pricing policy, Karma distribution and relative societal cost for an episode of training on a $n=3$ parallel-arc transportation network. We set $t_\mathrm{init}=20$ and $t_\mathrm{update}=5$.}
    \label{fig:MNL3Arc}
\end{figure}
Fig.~\ref{fig:MNL3Arc} depicts a learning episode of the toll-pricing controller of in total $t=100$ days. Convergence of $x(t)$ to $x^\star$ is achieved between $t=55$ and $t=65$, indicating the efficiency with which the optimal pricing policy is learned. Once converged, the average societal cost lies within 0.1\% of $C(x^\star)$. Throughout the episode, the societal cost $C(x)$ exceeds the cost of the uncontrolled scenario only at three individual time steps. These instances all fall under the initial pseudo-random policy. In fact, when we look at a 5-day moving average $C_\mathrm{MA}$, at no point is the cost of the uncontrolled scenario being exceeded.

\subsection{Non-stationary Dynamical Environment}
\label{sec:nonstationary}
A benefit of our data-driven method is its inherent ability of automatically coping with non-stationarities through the iterative updating instances of the controller. To display this automated learning process, we display the case where during a learning episode with two arcs, the road capacity parameters $\kappa$, and hence the dynamical system, are changed. More specifically, for $t\leq 80$ we set $\kappa=[\sfrac{1}{3},\sfrac{2}{3}]^\top$, whilst for $t>80$ we set $\kappa=[\sfrac{1}{2},\sfrac{2}{3}]^\top$. Furthermore, we set $d^0=[1,2]^\top$, $\alpha=0.15$ and $\beta=4$. Note that this setup entails two optimal- and uncontrolled equilibria $x^{\star,1}$, $x^{\mathrm{UC},2}$, $x^{\star,2}$ and $x^{\mathrm{UC},2}$. For the chosen parameters, these equilbria become $x^{\star,1} = [0.41,0.54 ]^\top$, $x^{\mathrm{UC},1} = [0.62,0.33]^\top$, $x^{\star,2} = [0.56,0.39 ]^\top$, $x^{\mathrm{UC},2} = [0.84,0.11]^\top$. 

\begin{figure}[tb]
    \centering
    \includegraphics[width=\columnwidth]{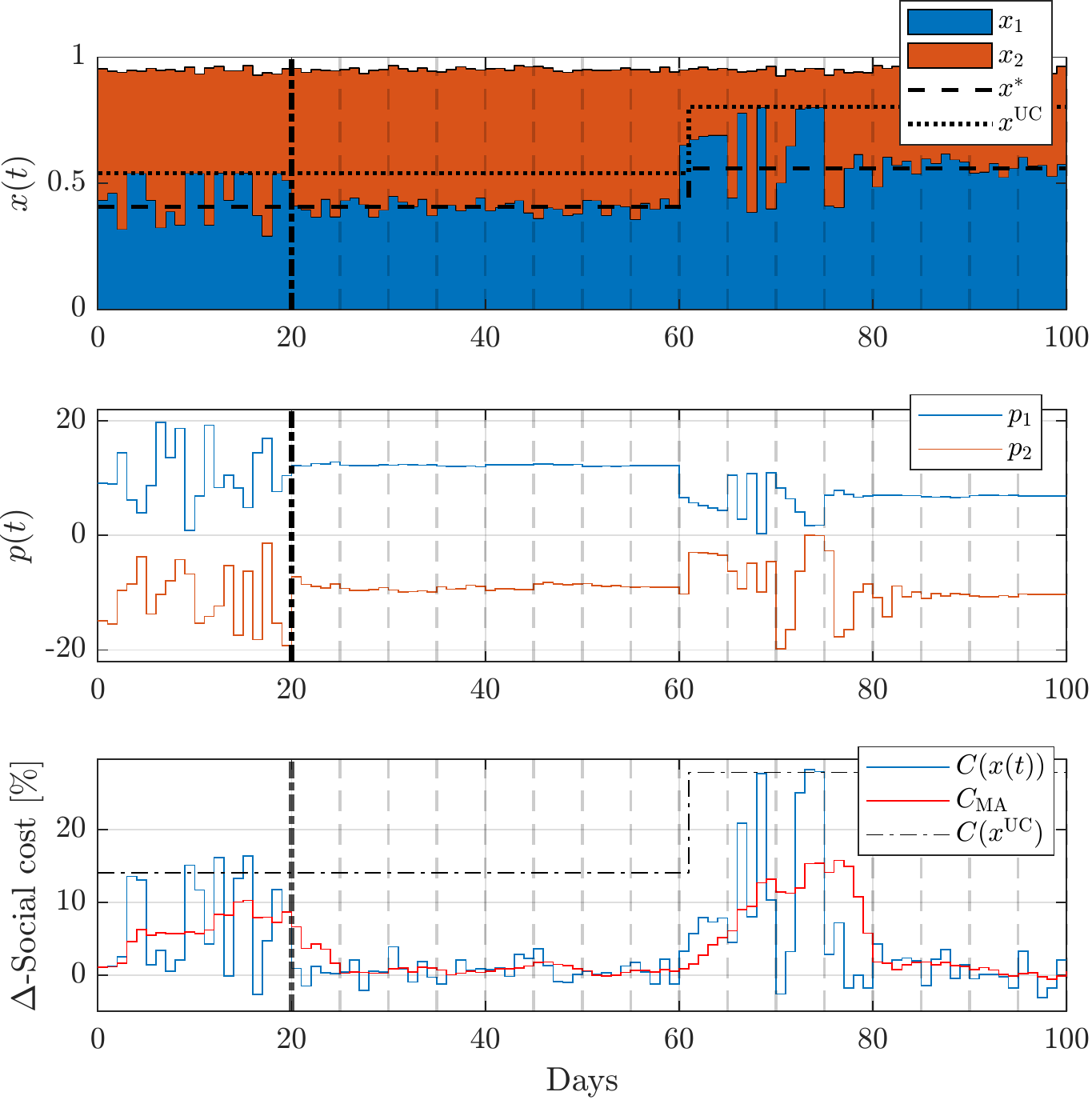}
    \caption{Numerical simulation of learning process in a non-stationary environment. We set \mbox{$k^i(0) \sim \mathcal{U}(50,100)$}, $t_{\mathrm{init}}=20$ and $t_{\mathrm{update}}=5$.}
    \label{fig:nonstationary}
\end{figure}

Fig.~\ref{fig:nonstationary} shows how convergence towards the first optimal road distribution $x^{\star,1}$ is obtained initially at $t=25$, after which the disturbance at $t=60$ forces the mesoscopic flows away from the new optimal distribution $x^{\star,2}$. After observing circa 20 days of this new environment, the toll-pricing controller is capable of once again reaching convergence at the new optimal road distribution $x^{\star,2}$. Despite phases of fluctuating prices, at no point throughout the episode does the 5-day moving average $C_\mathrm{MA}$ exceed the societal cost of the uncontrolled scenario $C(x^\mathrm{UC})$, indicating that users are always better off with the Karma policy in place.

\section{Conclusion}
\label{sec:Conclusion}
In this paper, we combined a data-driven pricing scheme with artificial currencies to route a group of selfish users in a system-optimal fashion within a repetitive-game setting. To this end, we developed a model-based reinforcement learning (RL) controller to autonomously devise an optimal Karma-based toll-pricing scheme. We showed that our method is effective in learning optimal pricing schemes for parallel-arc transportation networks consisting of more than two arcs. This indicates a step in scalability compared to previous analytic pricing schemes. Considering the assumed lack of a high-fidelity simulator, the significant efficiency satisfies one of the prerequisites for real-world implementation. Finally, we showed that our data-driven controller is capable of adjusting its pricing scheme accordingly, within a non-stationary dynamical environment, which removes the need of a local authority to periodically update its prices.

In future research, we would like to scale the transportation network further towards a more realistic size and complexity. Additionally, we wish to study the effect of smaller time windows, in order to optimize not only for spatial routing, but also temporal slots.
\bibliography{ifacconf}
\appendix
\end{document}